\documentclass{article}
\usepackage{spconf}
\usepackage{amsmath,amssymb}
\usepackage{graphicx}
\usepackage{booktabs}
\usepackage{cite}
\usepackage{tikz}
\usetikzlibrary{positioning,calc,fit,arrows.meta,decorations.pathreplacing}
\usepackage[colorlinks=true,linkcolor=black,citecolor=black,filecolor=black,urlcolor=blue]{hyperref}

\usepackage{stfloats}

\title{Load Disaggregation under Incomplete Inventory with Structured Unclaimed Demand}

\name{Saman Mostafavi \qquad Robert W. Cox}
\address{University of North Carolina at Charlotte\\
\texttt{\{smostafa,robert.cox\}@charlotte.edu}}

\AddToHookNext{shipout/foreground}{\put(0.5\paperwidth,-\paperheight){\raisebox{9mm}{\makebox(0,0)[b]{\parbox{\textwidth}{\centering\scriptsize Submitted to IEEE ICASSP 2027 and under review. This work has been submitted to the IEEE for possible publication. Copyright may be transferred without notice, after which this version may no longer be accessible.}}}}}

\begin{document}
\ninept

\maketitle

\begin{abstract}
An incomplete inventory creates a specific attribution ambiguity in load disaggregation:
unclaimed demand can offset changes in claimed loads without altering meter reconstruction.
We study how the allowed target and nuisance trajectories govern this ambiguity.
A bounded ordered Markov chain represents unclaimed demand, while an
unweighted acceptor removes commissioning-inconsistent paths from one
target without reweighting surviving paths; exact factorial MAP jointly
decodes all paths. At two fully instrumented commercial buildings,
restricting exterior-lighting support improves agreement on the
count of rooftop compressors although no compressor chain is constrained;
the gain depends on the nuisance representation. A synthetic experiment identifies the
boundary: the effect persists under support-preserving changes but
reverses when the true path violates commissioned support or the
commissioned level is miscalibrated. Attribution therefore depends on joint
target--nuisance path geometry, not reconstruction error alone.

\end{abstract}


\section{Introduction}
\label{sec:intro}

We study single-channel separation when a substantial, semantically
unclaimed component competes with the target sources: how do
restrictions on the admissible trajectories of one component alter
attribution to another? Hidden-state source decomposition provides the
direct methodological precedent \cite{varga1990,roweis2000},
and non-intrusive load monitoring (NILM) provides a real-world setting
with incomplete inventories and unusually strong source-level validation.
An aggregate meter observes the sum of all loads in a building,
although an operator may need only a few quantities from it.
NILM infers those quantities from aggregate power \cite{hart1992}.
The request treated here comes from a bank branch
with three rooftop units: how many of their compressors
are running, and whether exterior lighting is on.
Recent formulations leave loads outside the declared inventory unmodeled or in a residual term,
while the estimator is trained or specified for the declared appliances
\cite{kelly2015,zhang2018,sun2023diffnilm,yang2025mraff,petralia2025nilmformer}.

\begin{figure}[tb]
\centering
\includegraphics[width=\columnwidth]{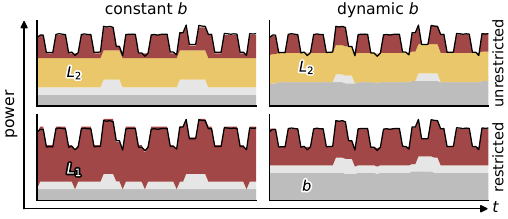}
\caption{One held-out Site-A afternoon decoded under the four combinations of 
constant/dynamic background and unrestricted/restricted support. $L_1$ is the counted rooftop chain, $L_2$ the
restricted exterior chain, $b$ unclaimed demand; truth has $L_1$
switching 18 times and $L_2$ off throughout. A constant $b$ leaves every
variation to the claimed loads, and $L_1$'s count is wrong at every
minute under either support. A dynamic $b$ with unrestricted support
switches $L_2$ on where its rule forbids it and undercounts $L_1$ by
one; restricted, it matches the breakers at every minute.}
\label{fig:geometry}
\end{figure}

At Site~A, instrumented at every breaker, unclaimed demand
averages $29\%$ of the aggregate, or
2.5 times the combined draw of the three rooftop units (RTUs).
Aggregate fit then does not determine attribution: changing a
target trajectory can be compensated by changing unclaimed demand, an exact
cancellation made precise in Equation~\eqref{eq:ambiguity} and shown on a
meter trace in \figurename~\ref{fig:geometry}.
The object to control is therefore not the estimator but the set of
target and nuisance trajectories considered admissible.

The interferer of HMM source decomposition, a chain with its own
dynamics and no semantics \cite{varga1990,roweis2000}, is the background
chain here. Within NILM, factorial models represent multiple appliance chains decoded jointly \cite{ghahramani1997,parson2012},
while learned estimators recover declared appliances directly from the aggregate \cite{zhang2018,yue2020},
as reviewed in \cite{schirmer2023review}.
AFAMAP assigns unclaimed demand a total-variation residual \cite{kolter2012};
related formulations address unknown loads, aggregate constraints,
and structured dynamics \cite{shaloudegi2016,zhong2014,makonin2016,quinn2009}. 
Constraints on latent paths have a long history in sequence inference \cite{roweis1999,mohri2002}.

Our contribution is to cross target-support and nuisance-model changes
as controlled interventions. Under incomplete inventory, a cost-neutral
restriction on one target's path support measurably changes attribution to a target it
never constrains, and the magnitude depends on what the nuisance model
leaves to correct.
An unweighted acceptor removes commissioning-inconsistent target paths
without changing any surviving path's cost \cite{mohri2002}; an ordered
Markov chain gives unclaimed demand a bounded, persistent explanation;
both are decoded jointly by exact factorial MAP \cite{ghahramani1997}.
Two operating buildings show the restriction improving a chain it never touches, and a
synthetic experiment shows where that ends.

\section{The Attribution Problem}
\label{sec:model}

A building is metered at one point. Some of its loads are of interest
and the rest are not. Write the measurement as a sum of two groups:
\begin{equation}
  y_t \;=\; \underbrace{\textstyle\sum_{k\in\mathcal K}\mu_k(z_{k,t})}
                       _{\text{claimed loads}}
        \;+\; \underbrace{b_t}_{\text{unclaimed demand}}
        \;+\; \varepsilon_t .
  \label{eq:split}
\end{equation}
Here $\mathcal K$ indexes the claimed loads, fixed at commissioning,
$z_{k,t}\in\{1,\dots,c_k\}$ is claimed load $k$'s state, $\mu_k$
its commissioned ladder, $b_t$ the demand outside that set, and
$\varepsilon_t\sim\mathcal N(0,\sigma^2)$.

The two groups are not separately identified. Let $z'$ be any
alternative trajectory for the target chains, and define
\begin{equation}
  b'_t = b_t + \sum_k \bigl[\mu_k(z_{k,t})-\mu_k(z'_{k,t})\bigr].
  \label{eq:ambiguity}
\end{equation}
Then $\sum_k\mu_k(z'_{k,t})+b'_t=\sum_k\mu_k(z_{k,t})+b_t$ at every
$t$. Whenever the constructed $b'$ is admissible, the explanations
have identical emission fit. The aggregate alone cannot separate
them. Support and path costs must do so. We measure how changes to the admissible set
move attribution, not whether the aggregate identifies a unique latent decomposition. At Site~A the four decoders of \figurename~\ref{fig:geometry} fit one
afternoon's meter trace within $0.1$--$0.3$\,kW while disagreeing on
the running count at almost every minute.

Equation~\eqref{eq:ambiguity} exposes two interventions: restrict the
target path set, or change which compensating paths the nuisance model
admits and how it prices them. A background too rigid leaves nuisance
variation to the claimed loads, whereas one too mobile can absorb target
steps. We manipulate these two levers separately.

\section{Controlling Admissible Explanations}
\label{sec:methodology}

\subsection{Cost-neutral support restriction of a target chain}
\label{sec:support}

Transition topology and finite-state constraints restrict latent paths
\cite{roweis1999,mohri2002}. A restriction is \emph{cost-neutral} when it
removes inadmissible trajectories without changing the costs of those
that remain, so that any change in the decode is attributable to
removal alone.

Let $\mathcal Z_k$ be chain $k$'s state paths, $A_k$ its transition
matrix and $\phi_k=-\log A_k$ its edge costs. Composing the chain with an unweighted acceptor from
commissioning \cite{mohri2002} retains a subset
$\mathcal S_k\subset\mathcal Z_k$ and leaves every surviving path at its
original cost,
\begin{equation}
  \textstyle\sum_t \phi'_k(z_{t-1},z_t) \;=\;
  \textstyle\sum_t \phi_k(z_{t-1},z_t)
  \qquad \forall\, z\in\mathcal S_k ,
  \label{eq:supp}
\end{equation}
with $\phi'_k$ the composed chain's edge costs; composition copies each
surviving edge's cost, so \eqref{eq:supp} holds by construction. The
composed chain is left unnormalized: its single global normalizer is
irrelevant to MAP, whereas renormalizing rows over the surviving edges
would reweight them and violate \eqref{eq:supp}.

At the deployed sites the acceptor encodes exterior lighting's
photocell rule, one ON interval wrapping the day boundary, equivalently
one OFF interval during daylight; over a day of $T$ samples,
\begin{equation}
  \mathcal S_{\mathrm{ext}}=\{\,1^{\ell_1}0^{\ell_0}1^{\ell_2}\;:\;
  \ell_i\ge 0,\ \ell_1+\ell_0+\ell_2=T\,\},
  \label{eq:acceptor}
\end{equation}
with $1$ the ON macro state. Empty phases retain all-dark and
one-switch days; only an additional daytime ON interval is excluded. More generally, the construction imposes a
fixed commissioned order of level blocks $w_1^{\ell_1}\cdots w_m^{\ell_m}$,
using one phase state per position in $w$; here $w=(1,0,1)$, so three
phase states replace two load states. Self-loops and forward edges retain the
original costs; every other edge receives $+\infty$.

\subsection{Structured unclaimed demand}

The background chain carries the unclaimed demand $b_t$ introduced
above. The labeled commissioning day fixes the support
endpoints $b_{\min}$ and $b_{\max}$; the grid spacing $\Delta$ and
motion scale $\tau_b$ are design constants fixed before held-out
decoding. At the fixed commissioned range, $B$ controls the grid
resolution through $\Delta=(b_{\max}-b_{\min})/(B-1)$, the smallest
representable change, while $\tau_b$ sets the prior scale of a
one-sample (one-minute) displacement. The background lies on a grid
$\mathcal G=\{b_{\min},\,b_{\min}{+}\Delta,\dots,b_{\max}\}$ of
$B=|\mathcal G|$ levels, with quadratic displacement cost
\begin{equation}
  \psi(b',b) = \frac{(b-b')^2}{2\tau_b^2}+\log Z(b'),
  \label{eq:walk}
\end{equation}
the negative log of a discrete Gaussian walk truncated to the grid, with
$Z(b')=\sum_{g\in\mathcal G}\exp[-(g-b')^2/(2\tau_b^2)]$. 

Discretization leaves a positive cost floor. Every nonzero increment is
at least $\Delta$, so a net change of magnitude $A$ incurs a displacement
cost of at least $A\Delta/2\tau_b^2$ however slowly it is taken. The
quadratic term further penalizes rapid motion. A continuous quadratic residual has
no such floor. At the deployed grid, absorbing a rooftop unit's
commissioned $3.4$\,kW running level costs the background at least
$24$ nats in displacement cost at any speed, about $120$ within
three samples and $360$ within one, while the unit's own chain takes
that step for about one nat. A target-sized abrupt step is far cheaper through the compressor
chain than through the background.

Unclaimed demand need not be dynamically homogeneous. The model
therefore permits an optional auxiliary chain $a_t$, with $c_A$ states,
emission ladder $\mu_A$ and transition cost $\phi_A$, to represent
structured switching demand while $b_t$ carries the diffuse remainder,
refining \eqref{eq:split} to
$y_t=\sum_k\mu_k(z_{k,t})+\mu_A(a_t)+b_t+\varepsilon_t$. Its inclusion
and parameters are fixed at commissioning and held identical across all
four decodes. Both enter one unclaimed estimate,
$\mu_A(\hat a_t)+\hat b_t$, and the auxiliary chain is never scored.

The nuisance intervention replaces the ordered chain with a single
level, fixed from the commissioning day and held constant over the
decoded day, with every target and auxiliary specification unchanged.
The level absorbs the commissioning offset. Every change over the day
is left to the claimed chains, the auxiliary chain and the noise.
Crossing the
constant/dynamic background choice with unrestricted/restricted exterior
support gives the four decoders evaluated below.

\subsection{Joint MAP and exact computation}

Decoding minimizes the joint path cost over the full product state
space: with $r_t=y_t-\sum_k\mu_k(z_{k,t})-\mu_A(a_t)-b_t$,
\begin{equation}
\begin{split}
  (\hat z,\hat a,\hat b) = \arg\min_{z\in\mathcal S,\,a,\,b}\;
  &\sum_t\frac{r_t^2}{2\sigma^2}
   +\sum_{k,t}\phi_k(z_{k,t-1},z_{k,t})\\
  &+\sum_t\bigl[\phi_A(a_{t-1},a_t)+\psi(b_{t-1},b_t)\bigr],
\end{split}
  \label{eq:map}
\end{equation}
with $\mathcal S$ the product of the claimed-chain path sets and
initial-state costs absorbed into the path costs. By \eqref{eq:supp},
the restriction enters only through $z\in\mathcal S$; $\phi_k$ remains
the original edge cost on every surviving path. Without an auxiliary
chain the $a$-terms are absent. With $c_A$ auxiliary states the joint
state count is $S=Bc_A\prod_k c_k$, $5400$ at Site~A unrestricted.
The chains couple only through the emission, so
the transition cost is separable and sequential min-plus elimination of
predecessor indices is exact.
The computational cost is $O(TS[B+c_A+\sum_k c_k])$, and decoding runs
in seconds per building-day on one CPU.
\section{Experiments}
\label{sec:results}

\subsection{Protocol and estimands}
\label{sec:setup}

\textbf{Design.} The experiment crosses two levers. The background is
either the ordered chain of \eqref{eq:walk} (dynamic $b$) or a single
commissioned level held constant over the day (constant $b$); the
exterior chain's support is either $\mathcal Z_{\mathrm{ext}}$
(unrestricted) or $\mathcal S_{\mathrm{ext}}$ of \eqref{eq:acceptor}
(restricted). Every other component is identical across the four arms,
including the claimed ladders, the noise scale and Site~A's auxiliary
chain, so each arm differs from its pair in one lever alone. The four
decodes run on the same days and are compared pairwise within a day. The
prespecified primary endpoint at each site is the restriction-induced
compressor-count gain under the dynamic background, the cross-chain
effect. Exterior agreement (the within-chain effect), water agreement,
mF1 and NDE$_R$ are secondary endpoints.

\textbf{Sites.} Two bank branches from a 2014 instrumentation study
\cite{mostafavi2019} are metered at every breaker at one-minute
resolution. Here $y_t$ is the breaker sum, so every term of
\eqref{eq:split}, $b_t$ included, has ground truth. Site~A (82
channels) claims three RTUs with $c_k=3$ (off, fan, fan{+}compressor),
which share one commissioned ladder and are therefore recovered only up
to label permutation, so the deliverable is their running count;
exterior and water have $c_k=2$, and one auxiliary chain has $c_A=2$
and is unscored. Site~B (107 breakers) claims four chains and no
auxiliary; its commissioned rule and cohort were fixed before either
support decode. The breaker channels fix the commissioning parameters,
are hidden during decoding, and are read again for scoring. Evaluation
cohorts are 178 days at Site~A (184 less six development dates) and 83
at Site~B. Count cohorts, fixed from truth alone to days with
compressor operation, comprise 120 Site-A days and 36 June--July Site-B
days. Outside that window, resistance heat shares the rooftop breakers.
The supervised comparators train on five sub-metered days outside their
respective cohorts, and the extended sequence-to-point arm on thirty
(\S\ref{sec:levers2}).

\textbf{Commissioning.} One labeled day per site (Site~A 1~May, Site~B
23~June 2014) instantiates \eqref{eq:split}: the ladders $\mu_k$ from
the aggregate alone, by step detection and magnitude clustering (the
aggregate recipe); $c_k$,
$A_k$, $\sigma$ and $[b_{\min},b_{\max}]$ from the labeled channels;
the constant background's level from the same day. The spacing
$\Delta$ and scale $\tau_b$ are design constants, fixed before the
held-out cohort was decoded and carried unchanged to Site~B
(\S\ref{sec:levers2} sweeps both). Deployed, A / B: $B{=}25$ / $52$
levels over $0.34$--$5.75$ / $3.33$--$14.82$\,kW, $\Delta{=}225$\,W,
$\tau_b{=}125$\,W, $\sigma{=}187$ / $251$\,W, $c_A{=}2$ at A only. On a
decoded day inference observes $y_{1:T}$ alone; the day's breaker
channels are read afterwards, for eligibility and scoring.

\textbf{Scoring.} Every method passes through one truth-normalized evaluation mapping from decoded
per-channel power $\hat p_{k,t}$ and the day's truth levels
$\mu^{(d)}_{k,\cdot}$:
\begin{equation}
\begin{split}
  \hat s_{k,t}&=\arg\min\nolimits_j
    \bigl|\hat p_{k,t}-\mu^{(d)}_{k,j}\bigr|,\\
  \hat K_t&=\textstyle\sum_{k\in\mathrm{RTU}}
  \mathbf 1\bigl\{\mu^{(d)}_{k,\hat s_{k,t}}-\mu^{(d)}_{k,1}
  \ge \theta_{\mathrm{c}}\bigr\},
\end{split}
  \label{eq:score}
\end{equation}
with $\theta_{\mathrm{c}}=1850$\,W between the largest commissioned fan
increment and the smallest compressor increment, and $K_t$ the metered
count. Count agreement is $T^{-1}\sum_t\mathbf 1\{\hat K_t=K_t\}$. NDE$_R$ is the day-mean of
$\sum_t(\hat r_t-r_t)^2/\sum_t r_t^2$ over the metered rooftop sum
$r_t=\sum_{k\in\mathrm{RTU}}\mu^{(d)}_{k,s_{k,t}}$ and its decoded
counterpart; mF1 averages macro-F1 over compressor-count classes,
exterior states and water states, never nominal RTU identities; every
table cell is the unweighted mean of per-day values. The mapping is
applied after decoding only, to every method and both support arms.
Intermediate-valued regressor outputs receive no partial credit: each
is assigned to the nearest truth rung; \S\ref{sec:levers2} also examines
the raw trajectories.

\textbf{Comparators.} AFAMAP \cite{kolter2012} over the predetermined
grid $\lambda_1\in\{0.3,1,2,3,5,10,30\}$, shown at its best count
weight, with a further arm swapping its total-variation residual for
the quadratic cost of \eqref{eq:walk} at the deployed $\tau_b$; an
exact per-appliance FHMM \cite{parson2012,ghahramani1997}, eight
deterministic EM initialisations selected on training likelihood;
sequence-to-point \cite{zhang2018}; BERT4NILM \cite{yue2020}, ported
unchanged; and combinatorial optimization \cite{hart1992}. Supervised
models train on per-appliance labels from the commissioning day and
next four usable days, reserved from both cohorts, with inventory, offset and noise scale fixed from
commissioning; the networks use seed~0. A thirty-day arm extends the
same rule to the commissioning day and next 29 days, with all methods
re-pooled over the 154 remaining days.

\textbf{Intervals.} Day-paired 95\% percentile bootstraps, 4000
replicates, moving-block resampled over the day-ordered paired
differences in blocks of 14 scored days at Site~A and by independent
days at Site~B, where 36 count days leave three blocks and that
interval is the widest of block lengths 1, 7 and 14.
Every block length gives the same sign, and differences are formed
before rounding.

\begin{table}[t]
\caption{The two levers crossed, and the paired compressor-count
contrasts they produce. Intervals: 95\% bootstraps over days, moving blocks of 14 at Site~A
and independent days at Site~B; the interaction is the
dynamic minus the constant-background gain.}
\label{tab:main}
\centering
\setlength{\tabcolsep}{1.2pt}
\begin{tabular*}{\columnwidth}{@{\extracolsep{\fill}}lcccccccc@{}}
\toprule
 & & \multicolumn{5}{c}{Site~A} & \multicolumn{2}{c}{Site~B} \\
\cmidrule(lr){3-7}\cmidrule(lr){8-9}
background & restr. & count & ext. & water & mF1 & NDE$_{\!R}$
 & count & ext. \\
\midrule
const. $b$ & no & .707 & .704 & .594 & .593 & 1.79 & .584 & .653 \\
const. $b$ & yes & .732 & .776 & .615 & .629 & 2.15 & .616 & .651 \\
dyn. $b$ & no & .910 & .953 & .939 & .864 & 0.40 & .736 & .838 \\
dyn. $b$ & yes & .948 & .977 & .952 & .897 & 0.38 & .866 & .888 \\
\midrule
majority floor & & .788 & .562 & .879 & --- & --- & .507 & .547 \\
\bottomrule
\end{tabular*}

\vspace{3pt}
\begin{tabular*}{\columnwidth}{@{\extracolsep{\fill}}lcc@{}}
\toprule
paired count contrast & Site~A & Site~B \\
\midrule
restr., const. $b$ & $+.025$~$[\text{-}.007,.054]$
 & $+.032$~$[.019,.045]$ \\
restr., dyn. $b$ & $+.038$~$[.008,.070]$
 & $+.130$~$[.091,.171]$ \\
interaction & $+.013$~$[\text{-}.015,.044]$
 & $+.098$~$[.061,.137]$ \\
\bottomrule
\end{tabular*}
\end{table}

\begin{table}[t]
\caption{Method comparison across per-appliance monitoring budgets. AFAMAP
oracle-best on the predetermined grid. Last row repeats
Table~\ref{tab:main}.}
\label{tab:comparators}
\centering
\setlength{\tabcolsep}{1.2pt}
\begin{tabular*}{\columnwidth}{@{\extracolsep{\fill}}lccccccc@{}}
\toprule
 & \multicolumn{5}{c}{Site~A} & \multicolumn{2}{c}{Site~B} \\
\cmidrule(lr){2-6}\cmidrule(lr){7-8}
method & count & ext. & water & mF1 & NDE$_{\!R}$ & count & ext. \\
\midrule
\multicolumn{8}{@{}l}{\itshape days of per-appliance monitoring: five} \\
seq2point \cite{zhang2018} & .822 & .967 & .881 & .782 & 0.53 & .795 & .851 \\
BERT4NILM \cite{yue2020} & .823 & .901 & .879 & .667 & 0.70 & .379 & .785 \\
supervised FHMM \cite{parson2012,ghahramani1997} & .813 & .792 & .891 & .778 & 0.95 & .680 & .804 \\
\midrule
\multicolumn{8}{@{}l}{\itshape one sub-metered commissioning day, then aggregate only} \\
AFAMAP \cite{kolter2012} $\lambda_1{=}0.3$ & .867 & .678 & .952 & .741 & 0.42 & .726
 & .635 \\
AFAMAP, quadratic & .845 & .746 & .953 & .775 & 0.82 & .705 & .661 \\
comb.\ opt. \cite{hart1992} & .639 & .606 & .409 & .504 & 3.32 & .409 & .600 \\
\midrule
ours, deployed & .948 & .977 & .952 & .897 & 0.38 & .866 & .888 \\
\bottomrule
\end{tabular*}
\end{table}

\subsection{Cross-chain effect of the restriction}
\label{sec:supportresult}

The background is the dominant lever (Table~\ref{tab:main}): the
dynamic chain raises count agreement at both sites, and at Site~A it
brings the decoded unclaimed demand to within $0.53$\,kW RMSE and
$0.05$\,kW bias of its metered truth, from $1.50$ and $-0.78$\,kW under
the constant level. The restriction then acts on what that model
leaves.

Under the dynamic background, restricting only exterior lighting raises
the prespecified compressor-count agreement at both sites without
constraining a compressor chain. Both paired intervals exclude zero
(Table~\ref{tab:main}).\footnote{Scoring the high-power state on all
83 Site-B days, the restriction moves that count $.790\to.862$,
$+.073$ $[+.031,+.119]$, against AFAMAP's $.781$.} Exterior agreement,
the within-chain effect, rises by $+.025$ $[+.001,+.050]$ at Site~A and
from $.838$ to $.888$ $[-.016,+.114]$ at Site~B. The
background-by-restriction interaction is resolved at Site~B and not at
Site~A (Table~\ref{tab:main}): a background free to move is where a
target-sized step otherwise lands.

\subsection{Comparator failure modes and robustness}
\label{sec:levers2}

\begin{figure}[t]
\centering
\includegraphics[width=\columnwidth]{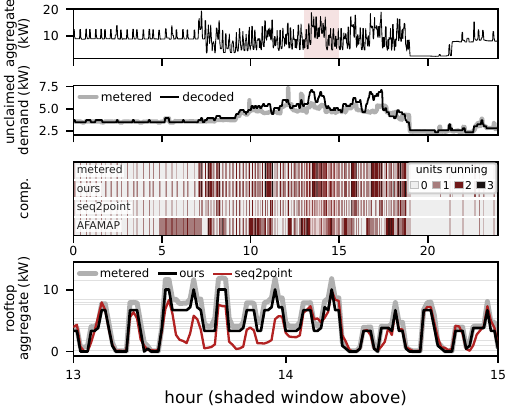}
\caption{Failure anatomy on a Site-A development day, 9 June 2014.
Sequence-to-point predicts between commissioned rooftop levels and
undercounts multi-unit peaks. AFAMAP's count persists across low-count
gaps yet misses high concurrency. The deployed decoder tracks the
metered count while assigning continuous variation to unclaimed
demand. Bottom: rooftop outputs from 13--15\,h.}
\label{fig:day}
\end{figure}

Table~\ref{tab:comparators} conceals distinct failure modes
(\figurename~\ref{fig:day}). Sequence-to-point follows event
timing but predicts between commissioned rooftop levels. Under the
nearest-rung mapping, these deficits accumulate during
multi-unit operation, most at three-unit peaks. AFAMAP's
discrete path fails differently on this day: it persists
across low-count gaps yet misses high-concurrency peaks. Pooled
three-unit agreement is $.07$ for sequence-to-point and $.14$ for
AFAMAP, against $.98$ for the deployed decoder. With thirty training
days, the deployed decoder retains a $+.095$ count-agreement advantage
$[+.001,+.185]$, while exterior agreement is tied.

AFAMAP's total-variation residual plays the role of unclaimed demand.\addtocounter{footnote}{1}\footnotetext{Generator, configuration and decoder:
\url{https://github.com/shantropy/icassp2027-incomplete-inventory-nilm}}\addtocounter{footnote}{-1}
Sweeping its penalty traces a count--exterior tradeoff
(\figurename~\ref{fig:mech}a), and neither applying the same acceptor
nor replacing the residual with the quadratic walk reaches the
deployed point. For $B\ge25$ and $\tau_b$ from
$60$ to $250$\,W, restricted-arm count agreement stays within $.03$
of the deployed cell (\figurename~\ref{fig:mech}b). A $\pm20\%$
compressor-ladder error leaves that count between $.927$ and
$.947$, and breaker-derived ladders move it to $.946$.

\subsection{Synthetic boundary test}
\label{sec:synth}

\begin{figure}[h]
\centering
\includegraphics[width=\columnwidth]{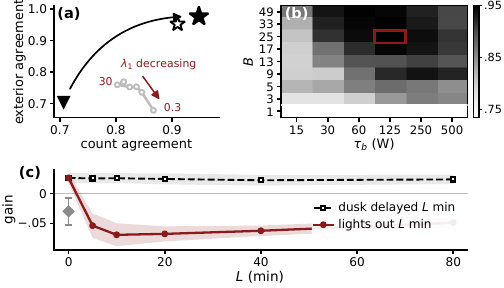}
\caption{(a) Comparator geometry. Varying AFAMAP's residual weight
$\lambda_1$ (grey, 30 to 0.3) traces a count--exterior tradeoff. Triangle, constant $b$.
Open and filled stars, dynamic $b$ without and with the restriction.
(b) Robustness. Restricted-arm count agreement over $(B,\tau_b)$.
Outline, deployed point. $B{=}1$, constant $b$.
(c) Synthetic boundary. A delayed dusk preserves the commissioned
support, a night outage violates it. Bands, 95\% intervals over 50
days. Diamond, the aggregate-recipe level, $14\%$ high.}
\label{fig:mech}
\end{figure}

Whether the gain comes from the specific support
$\mathcal S_{\mathrm{ext}}$ or from any shrinkage of the path set is
tested on a synthetic site\footnotemark:
three interchangeable rooftop units, a photocell exterior chain and
unclaimed demand built from plateaus, steps and ramps, not from the
walk of \eqref{eq:walk}, commissioned from
one labeled day and decoded on fifty days from
the aggregate alone. Under the protocol frozen
before decoding, with ladders from the aggregate recipe, the
restriction changes compressor-count agreement by $-.030$ $[-.053,-.008]$ under the
dynamic background: a fan start coincides with dusk that
day, the recipe places the exterior level $14\%$ high,
and the retained lighting path costs more than a phantom compressor
two levels below. With the exterior level read from the labeled day
instead, a sensitivity run after that result, the restriction raises
compressor-count agreement by $+.026$ under the dynamic background and
$+.073$ $[+.042,+.104]$ under the constant one. Here the constant
background produces the larger gain. This reverses Site~B: greater
background mobility does not uniformly strengthen the restriction. The gain has the
predicted boundary (\figurename~\ref{fig:mech}c): a dusk
delayed by $L$ minutes stays inside $\mathcal S_{\mathrm{ext}}$ and the
gain holds, while an $L$-minute night outage leaves it and reverses
the gain. Because surviving-path costs are unchanged, the sign reverses when
the true path leaves the support or level error makes the retained
path costlier than a competing attribution. The gain belongs
to the commissioned support, not restriction alone.

\section{Conclusion}
\label{sec:conclusion}

Under an incomplete inventory, equal aggregate reconstructions can
assign power to different targets, so attribution is governed by the
trajectories the decoder admits. Restricting exterior-lighting support
at unchanged path cost raised rooftop-compressor count agreement at two
sites without constraining a compressor chain. The gain depended on
what the nuisance model left to correct. On a synthetic site the
gain held under support-preserving perturbations and reversed when the
true trajectory violated the commissioned support or the commissioned
level was wrong. The design object is the joint target--nuisance path
geometry, not reconstruction error alone.

\clearpage
\bibliographystyle{IEEEbib}
\bibliography{refs}

\begin{thebibliography}{10}

\bibitem{varga1990}
A.~P. Varga and R.~K. Moore,
\newblock ``Hidden {M}arkov model decomposition of speech and noise,''
\newblock in {\em Proc. IEEE Int. Conf. Acoustics, Speech and Signal Processing
  (ICASSP)}, 1990, pp. 845--848.

\bibitem{roweis2000}
S.~Roweis,
\newblock ``One microphone source separation,''
\newblock in {\em Advances in Neural Information Processing Systems 13}, 2000,
  pp. 793--799.

\bibitem{hart1992}
George~W. Hart,
\newblock ``Nonintrusive appliance load monitoring,''
\newblock {\em Proceedings of the IEEE}, vol. 80, no. 12, pp. 1870--1891, 1992.

\bibitem{kelly2015}
Jack Kelly and William Knottenbelt,
\newblock ``Neural {NILM}: Deep neural networks applied to energy
  disaggregation,''
\newblock in {\em Proc. ACM BuildSys}, 2015.

\bibitem{zhang2018}
Chaoyun Zhang, Mingjun Zhong, Zongzuo Wang, Nigel Goddard, and Charles Sutton,
\newblock ``Sequence-to-point learning with neural networks for non-intrusive
  load monitoring,''
\newblock in {\em Proc. AAAI}, 2018.

\bibitem{sun2023diffnilm}
Ruichen Sun, Kun Dong, and Jianfeng Zhao,
\newblock ``{DiffNILM}: A novel framework for non-intrusive load monitoring
  based on the conditional diffusion model,''
\newblock {\em Sensors}, vol. 23, no. 7, pp. 3540, 2023.

\bibitem{yang2025mraff}
Lingxi Yang, Meijun Sun, Haowei Ran, Yipu Liu, Yan Zhou, and Zheng Wang,
\newblock ``Multi-scale re-weighted attention feature fusion for non-intrusive
  load monitoring,''
\newblock in {\em Proc. IEEE Int. Conf. Acoustics, Speech and Signal Processing
  (ICASSP)}, 2025.

\bibitem{petralia2025nilmformer}
Adrien Petralia, Philippe Charpentier, Youssef Kadhi, and Themis Palpanas,
\newblock ``{NILMFormer}: Non-intrusive load monitoring that accounts for
  non-stationarity,''
\newblock in {\em Proc. ACM SIGKDD Conf. Knowledge Discovery and Data Mining},
  2025.

\bibitem{ghahramani1997}
Zoubin Ghahramani and Michael~I. Jordan,
\newblock ``Factorial hidden {M}arkov models,''
\newblock {\em Machine Learning}, vol. 29, pp. 245--273, 1997.

\bibitem{parson2012}
Oliver Parson, Siddhartha Ghosh, Mark Weal, and Alex Rogers,
\newblock ``Non-intrusive load monitoring using prior models of general
  appliance types,''
\newblock in {\em Proc. AAAI}, 2012.

\bibitem{yue2020}
Zhenrui Yue, Camilo~Requena Witzig, Daniel Jorde, and Hans-Arno Jacobsen,
\newblock ``{BERT4NILM}: A bidirectional transformer model for non-intrusive
  load monitoring,''
\newblock in {\em Proc. NILM Workshop}, 2020.

\bibitem{schirmer2023review}
Pascal~A. Schirmer and Iosif Mporas,
\newblock ``Non-intrusive load monitoring: A review,''
\newblock {\em IEEE Transactions on Smart Grid}, vol. 14, no. 1, pp. 769--784,
  2023.

\bibitem{kolter2012}
J.~Zico Kolter and Tommi Jaakkola,
\newblock ``Approximate inference in additive factorial {HMM}s with application
  to energy disaggregation,''
\newblock in {\em Proc. AISTATS}, 2012, pp. 1472--1482.

\bibitem{shaloudegi2016}
Kiarash Shaloudegi, Andr{\'a}s Gy{\"o}rgy, Csaba Szepesv{\'a}ri, and Wilsun Xu,
\newblock ``{SDP} relaxation with randomized rounding for energy
  disaggregation,''
\newblock in {\em Advances in Neural Information Processing Systems}, 2016.

\bibitem{zhong2014}
Mingjun Zhong, Nigel Goddard, and Charles Sutton,
\newblock ``Signal aggregate constraints in additive factorial {HMM}s, with
  application to energy disaggregation,''
\newblock in {\em Advances in Neural Information Processing Systems}, 2014.

\bibitem{makonin2016}
Stephen Makonin, Fred Popowich, Ivan~V. Baji{\'c}, Bob Gill, and Lyn Bartram,
\newblock ``Exploiting {HMM} sparsity to perform online real-time nonintrusive
  load monitoring,''
\newblock {\em IEEE Transactions on Smart Grid}, vol. 7, no. 6, pp. 2575--2585,
  2016.

\bibitem{quinn2009}
John~A. Quinn, Christopher K.~I. Williams, and Neil McIntosh,
\newblock ``Factorial switching linear dynamical systems applied to
  physiological condition monitoring,''
\newblock {\em IEEE Transactions on Pattern Analysis and Machine Intelligence},
  vol. 31, no. 9, pp. 1537--1551, 2009.

\bibitem{roweis1999}
S.~T. Roweis,
\newblock ``Constrained hidden {M}arkov models,''
\newblock in {\em Advances in Neural Information Processing Systems 12}, 1999,
  pp. 782--788.

\bibitem{mohri2002}
M.~Mohri, F.~Pereira, and M.~Riley,
\newblock ``Weighted finite-state transducers in speech recognition,''
\newblock {\em Computer Speech \& Language}, vol. 16, no. 1, pp. 69--88, 2002.

\bibitem{mostafavi2019}
Saman Mostafavi,
\newblock {\em Minimally Supervised Low-Frequency Methods for Monitoring Major
  Energy Consumers in Buildings},
\newblock Ph.D. thesis, The University of North Carolina at Charlotte, 2019.

\end{thebibliography}

\end{document}